\documentclass[aps,prl,twocolumn,superscriptaddress,nofootinbib,showpacs,showkeys,preprintnumbers]{revtex4-1}
\usepackage{graphicx}  
\usepackage{dcolumn}   
\usepackage{bm}        
\usepackage{amssymb}   
\usepackage{lineno}
\usepackage{draftcopy}

\usepackage{color}
\definecolor{dgreen}{cmyk}{1.,0.,1.,0.2}        
\definecolor{orange}{cmyk}{0.,0.353,1.,0.}    

\usepackage[bookmarks]{hyperref}

\def\snn{\mbox{$\sqrt{s_{_{\rm NN}}}$}}

\newcommand{ \be }{\begin{eqnarray}}
\newcommand{ \ee }{\end{eqnarray}}

\newcommand{ \eps }{\varepsilon}
\newcommand{ \mean }[1]{\left\langle #1 \right\rangle}   

\newcommand{ \psin }{\Psi_{n}}

\newcommand{\pt}{p_T}

\begin{document}




\title{Ultra-relativistic nuclear collisions: where the spectators flow?}

\author{Sergei A. Voloshin and Takafumi Niida}
\affiliation{Wayne State University, 666 W. Hancock, Detroit, MI 48201}

\begin{abstract}
In high energy heavy ion collisions, the directed flow of particles is 
conventionally measured with respect to that of the projectile
spectators, which is defined as positive $x$ direction. But it is
not known  if the spectators deflect in the ``outward'' direction or
``inward'' -- toward the center line of the collision. In this Letter we
discuss how the measurements of the directed flow at mid-rapidity,
especially in asymmetric collision such as Cu+Au, can be used
to answer this question. We show that the existing data strongly favor
the case that the spectators, in the ultrarelativistic collisions,
on average deflect outwards.
    
\end{abstract}


\maketitle

In an ultrarelativistic nuclear collision only part of all nucleons
from the colliding nuclei experience a truly inelastic collision. Some
of nucleons, called spectators, stay mostly intact (or might
experience a transition to an excited state). Nevertheless, those
nucleons do experience a nonzero momentum transfer and deflect from
the original nucleus trajectory. The direction of such projectile
nucleon (``spectator'') deflection is conventionally taken as a
positive $x$ direction in the description of any anisotropic particle
production (anisotropic flow~\cite{Voloshin:2008dg}). At the same
time, while this direction has been measured experimentally at very
low collision energies, nothing is known on which direction the
spectators really deflect at high energies -- toward the center of the
collision, or outwards. Note that this question is not of a pure
``academic'' interest, it is intimately related to understanding of
the nucleon wave function in the nucleus, as well as momentum
distribution of the nucleons confined in a
nucleus~\cite{Alvioli:2010yk}. It is also important for the
interpretation of the anisotropic flow measurements. In particular,
the knowledge of the spectator flow is requited for determination of
the direction of the magnetic field created in the collision as well
as the system orbital momentum. The latter, for example, is needed for
the measurements of the so-called global
polarization~\cite{Liang:2004ph,Voloshin:2004ha,Abelev:2007zk}.

The only (known to authors) direct determination of the spectator
nucleons deflection direction was performed at the energies
$E/A\sim$100~MeV by measuring of the polarization of emitted
photons~\cite{Lemmon:1999}.  It was observed (see
also~\cite{Krofcheck:1989nda,Ogilvie:1990zz}) that around this energy
the direction of the deflection direction changes from the ``in-ward''
(due to attractive potential at lower energies) to the ``out-ward'' at
higher energies. No similar measurements was performed at higher
collision energies. Theoretically, this question is also not well
understood. As recently has been shown in~\cite{Alvioli:2010yk}, the
direction of the spectator deflection is likely dependent on the
nucleon transverse momentum. These calculations show that at
relatively large transverse momentum (more than $\sim$200~MeV) the
nucleons are likely deflected inwards, while at low transverse
momentum they might deflect outwards. One reason for the latter might
be the Coulomb interaction (repulsion) of the spectator protons.

\begin{figure}
\includegraphics[keepaspectratio, width=1.\columnwidth]{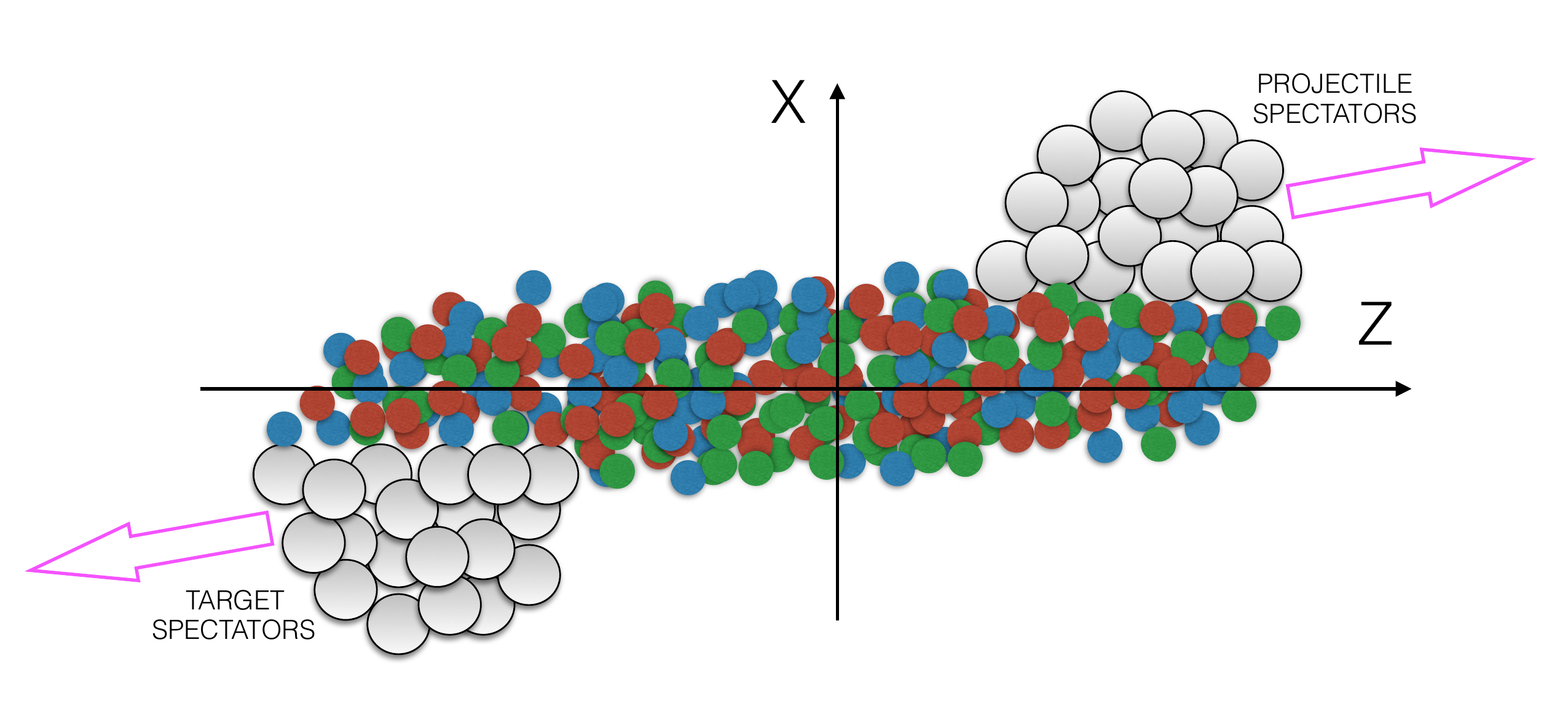}
  \caption{
    Schematic view of the collision. Arrows indicate the direction of
    the spectator flow; in the figure -- ``outward'' from the center-line.}
  \label{fig:coll}
\end{figure}

In this article we show how the study of the charge particle directed
flow at midrapidity measured relative to the spectator deflection
direction (directed flow) can help to answer the question of which
direction the spectators are deflected on average. We do not
distinguish between low and high $\pt$ spectators in this study,
though in principle this question can be studied experimentally.

The main idea of our approach is based on the observation that in the
case of asymmetric initial density distribution in the system, the
high(er) transverse momentum particles on average are flowing/emitted
in the direction of the largest density gradient, while the lower
$\pt$ particles flow in the opposite
direction~\cite{Heinz:2004et,Teaney:2010vd}. If the mean transverse
momentum of all particles is zero (e.g at midrapidity region in
symmetric collisions) then the average, integrated over all
transverse momenta, directed flow is in the same direction as that of
low $\pt$ particles.

Then the strategy in the establishing the direction of the spectator
flow becomes straight-forward. First, one has to measure the directed
flow of particles at midrapidity with respect to the spectator
deflection. 
Comparing that to the initial density gradients calculated
relative to the position of spectators, one can determine the
direction of spectator flow.  
The direction of the highest density
gradient in the system has to be determined with the help of a model,
but this appears to be a very robust procedure, as this direction
depends mostly on the distribution of the matter inside the
nucleus. As we argue below, there is no real model
dependence/ambiguity here.  In asymmetric collisions, such as Cu+Au,
the direction of the density gradient can be established unambiguously
on average, over all events. In symmetric collisions, e.g. Au+Au at
RHIC or Pb+Pb at LHC, one has to account for the fluctuation nature of
the density distribution and look for the density gradients relative
to the position of the spectators.

To quantify the anisotropic flow we use a standard Fourier
decomposition of the azimuthal particle distribution with respect to
the $n$-th harmonic symmetry
planes~\cite{Voloshin:1994mz,Alver:2010gr}:
\begin{equation}
E\frac{d^3 N}{d^3 p} = \frac{1}{2\pi} \frac{d^2 N}{p_{T} dp_{T}
  dy}\! \left(1\!+\!\sum_{n=1}^{\infty}2v_n \cos[n\!\left(
\phi\!-\!\Psi_n \right) ] \!\right), 
\label{eqFourier}
\end{equation}
where $v_n$ is the $n$-th harmonic flow coefficient and $\psin$ is
the $n$-th harmonic symmetry plane determined by the initial geometry
of the system (as given by the participant nucleon distribution, see below).
According to model calculations (see~\cite{Noronha-Hostler:2015dbi} and
references therein)
the event-by-event fluctuations in anisotropic flow closely
follow the fluctuations in the corresponding eccentricities of the
initial density distribution.  Following~\cite{Teaney:2010vd}, for the
latter we use the definition, for $n\ge 2$:
\be
\eps_{n,x}=\eps_{n}\cos(n\Psi_n)=-\mean{r^n \cos(n \phi)}/\mean{r^n}
 \\
\eps_{n,y}=\eps_{n}\sin(n\Psi_n)=-\mean{r^n \sin(n\phi)}/\mean{r^n}
\ee 
and for $n=1$ (most important for this study)
\be
\eps_{13,x}=\eps_{13}\cos(\Psi_{13})=-\mean{r^3 \cos(\phi)}/\mean{r^3}
 \\
\eps_{13,y}=\eps_{13}\sin(\Psi_{13})=-\mean{r^3 \sin(\phi)}/\mean{r^3}
\ee 
where $\eps_{n}=\sqrt{\eps_{n,x}^2+\eps_{n,y}^2}$ is the so-called
{\em participant} eccentricity~\cite{Alver:2006wh}. In our Monte-Carlo
model, in calculations of the average quantities in eccentricity
definitions we weight with the number of participating nucleons (those
undergoing inelastic collision).  For the nucleon distribution in the
nuclei we use the Woods-Saxon density distribution with standard
parameters (for the exact values see~\cite{Voloshin:2010ut}); the
inelastic nucleon-nucleon cross section is taken to be 42~mb for
calculations of at $\snn=200$~GeV (Cu+Au collisions discussed below) and
64~mb for $\snn=2.76$~TeV (Pb+Pb collisions).  In our model
calculations we chose the positive ``x'' direction to point along the
impact parameter vector, and assume that the spectators deflect in the
``outwards'' direction (target spectators flow in the impact parameter
vector direction, as indicated in Fig.~\ref{fig:coll}), and then check
if this agrees with the experimental observations.

There exist several measurements of directed flow at midrapidity
relative to the spectator nucleons in Au+Au and Cu+Cu collisions at
RHIC. Unfortunately, all those measurements reported only rapidity odd
component of the directed flow, that is not suitable for our
discussion, as in symmetric collision this component is exactly zero
at midrapidity. Rapidity even component, not zero at midrapidity even in
symmetric collisions due to fluctuations in initial density
distribution, has been measured only in Pb+Pb collisions at LHC by
ALICE Collaboration~\cite{Abelev:2013cva}. We will analyze these
measurements below first, and then discuss less ambiguous directed
flow measurements in asymmetric Cu+Au collisions at $\snn=200$~GeV by
PHENIX~\cite{Adare:2015cpn} and STAR~\cite{Niida:2016khm}
Collaborations.

\begin{figure}
\includegraphics[keepaspectratio, width=0.95\columnwidth]{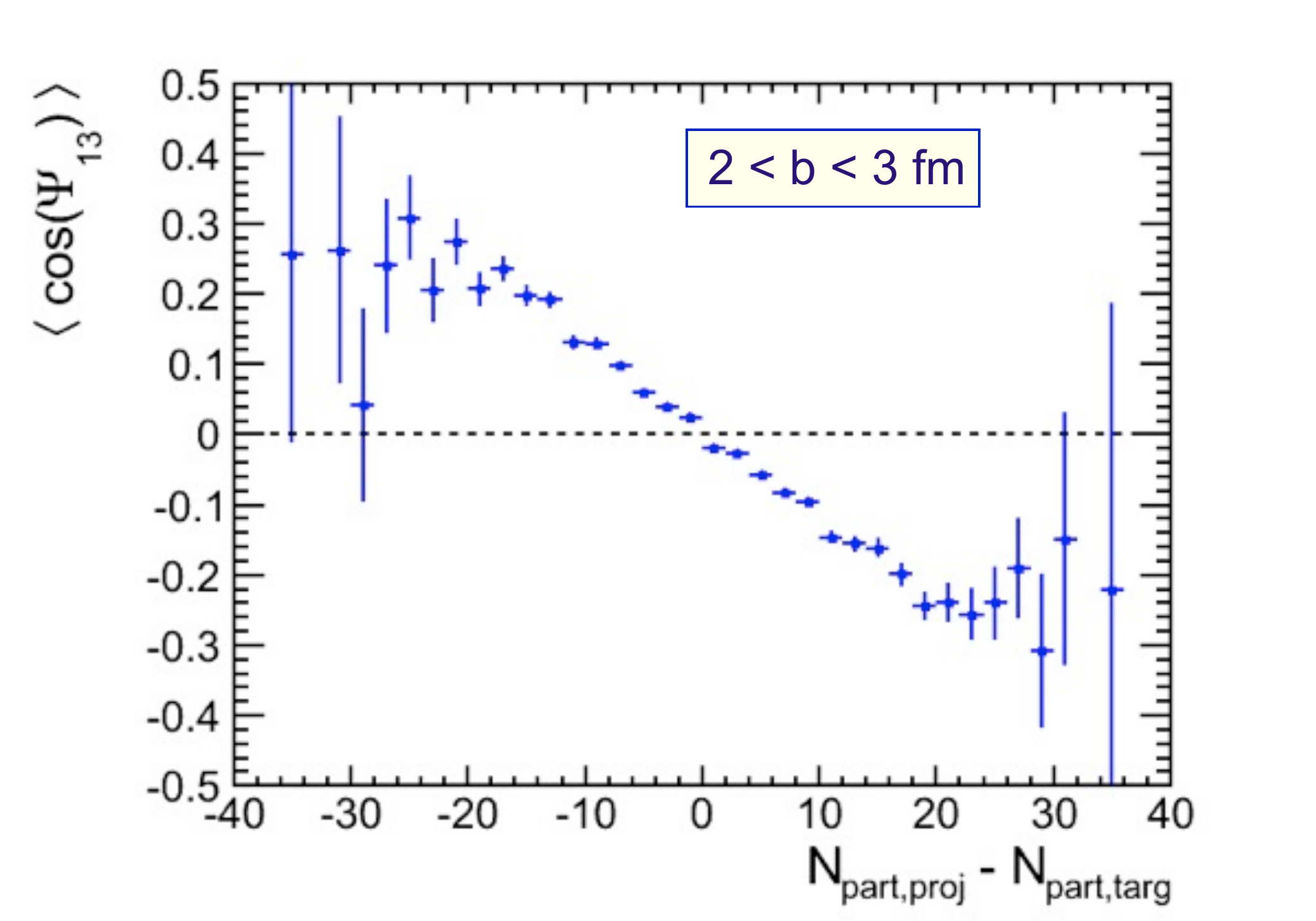}
  \caption{
$\mean{\cos(\Psi_{13})}$ as function of the difference in number of
    target and projectile nucleon participants in Pb+Pb collision in the
    impact parameter range $2<b<3$~fm.   
}
  \label{fig:psi13npart}
\end{figure}

In symmetric nuclear collisions, such as Pb+Pb, the directed flow at
midrapidity due to density fluctuations, if measured relative to the
projectile spectator flow, can be non-zero only due to decorrelation
in the flow directions of target and projectile spectators (and
corresponding geometry) or fluctuations in the relative reaction plane
resolutions due to fluctuations in the number of the spectators. We test
the latter by calculating the directed flow at midrapidity,
$\cos(\Psi_{13})$, as a function of the difference in the number of
projectile and target participants. An example of such calculations
for the impact parameter range $2<b<3$~fm is shown in
Fig.~\ref{fig:psi13npart}. From that plot it follows that in the case
of the smaller number of projectile participants
$\mean{\cos(\Psi_{13})}>0$ and the average directed flow would be
negative. The smaller number of participants corresponds to the larger
number of spectators that have to lead to better event plane
resolution and thus dominate the measurements. Having in mind that the
measurements~\cite{Abelev:2013cva} indicate {\em negative} rapidity
even component of the directed flow one has to conclude that the flow
of spectators must be ``outward'' (as assumed in the model). This
reasoning one can check with direct measurement of flow as a function
of the difference in number of spectators (e.g. as measured by zero
degree calorimeters). Unfortunately at present there is no such
results published.

\begin{figure}
\includegraphics[keepaspectratio, width=0.95\columnwidth]{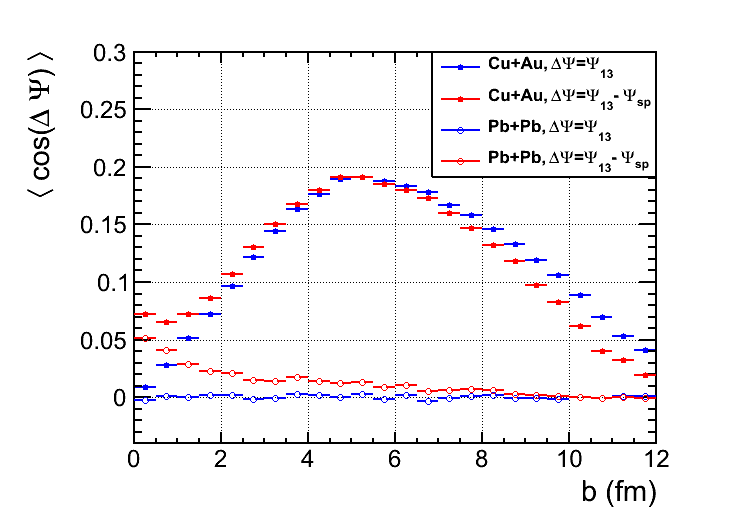}
  \caption{ $\mean{\cos(\Psi_{13}-\Psi_{sp})}$ and $\mean{\cos(\Psi_{13})}$
as function of the impact parameter for Pb+Pb collisions at
$\snn=2.76$~TeV (open markers) and Cu+Au collisions at 200~GeV (filled
markers). In Cu+Au collisions the Au nucleus is defined as the projectile;
$\Psi_{sp}$ is calculated using Au spectators.
}
  \label{fig:cosPsi13}
\end{figure}

The effect of the projectile and target spectator flow direction
decorrelation, and the correlations of the corresponding directions
with the direction of the density gradient at midrapidity,
 can be studied as follows.
Let us assume that the direction of the spectator flow is
along the line between the center of the nucleus and the ``center of
gravity'' of the projectile spectators in the transverse plane.  We
denote the corresponding angle $\Psi_{sp}$.  We calculate the
correlation of that angle with $\Psi_{13}$, indicative of the
direction of the (participant) density gradients that determined
directed flow at midrapidity. The results of these calculations for
Pb+Pb collision are shown in Fig.~\ref{fig:cosPsi13} by open red
markers. One can clearly see a positive correlations, which again
would lead to a conclusion that an average the flow at midrapidity
should be negative (recall that
on average the directed flow is in the opposite direction to
$\Psi_{13})$). Blue open points in Fig.~\ref{fig:cosPsi13} show the results for
$\mean{\cos(\Psi_{13})}$ and are consistent with zero as expected for
symmetric nuclear collisions.

While the discussion above about directed flow at midrapidity in
symmetric collisions is based on rather subtle details of the
treatment/modeling of the fluctuations in the initial density
distributions, in the asymmetric collisions, such as Cu+Au, the
direction of the density gradient practically is insensitive to the
fluctuations.  In this case, the line of arguments and the conclusion
become totally unambiguous. In the calculations discussed below we
treat Au nucleus as the projectile, and Au spectators are used 
in calculations of the angle $\Psi_{sp}$. 

\begin{figure}
\includegraphics[keepaspectratio, width=0.9\columnwidth]{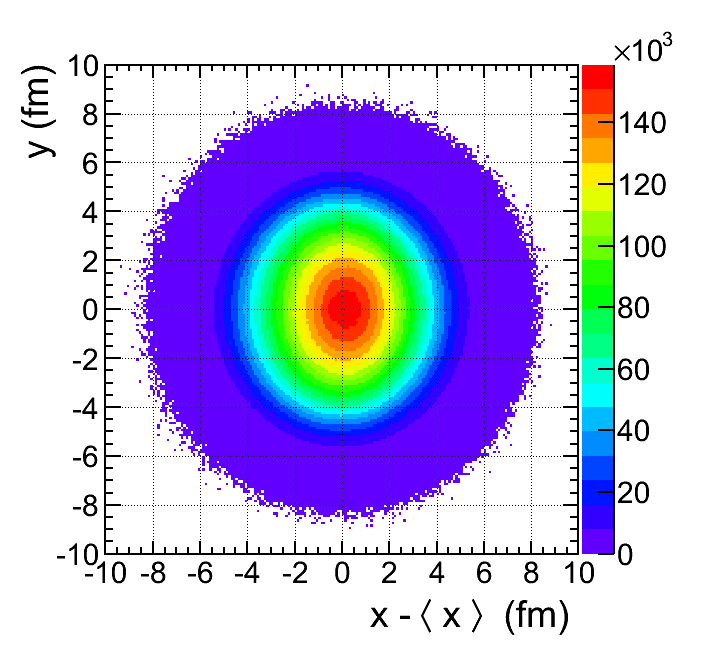}
  \caption{
    Participant distribution in Cu+Au collisions in the impact
    parameter range $2<b<3$~fm. Positive $x$ direction is toward the
  Au nucleus.}
  \label{fig:d2xy}
\end{figure}

Figure~\ref{fig:d2xy} presents the nucleon participant distributing in
Cu+Au collisions in the impact parameter range $2<b<3$~fm. The
distribution looks rather symmetric, but a more detail study 
indicates that the density gradient is larger in the positive ``x''
direction. This is clearly seen in Fig.~\ref{fig:cosPsi13} (filled
blue points). The effect of the density fluctuations and the corresponding
correlations between the density gradients and the position of
spectators (shown by red points) is rather insignificant in this case
unless one considers very central collisions. In peripheral collision
we observe that the red points are slightly below the blue points,
which can be explained by the decorrelations of the direction of
spectator flow relative to the reaction plane determined by the impact
parameter.    

The measurements of directed flow at midrapidity in Cu+Au
collisions~\cite{Adare:2015cpn,Niida:2016khm} show that charge
particles at midrapidity on average flow in the opposite direction to
that of the projectile spectators. Thus, once again, we are to
conclude that on the average the spectators flow ``outward'' from the
collision center.  We note that the experimental values of the mean
$v_1$ in Cu+Au collisions is about an order of magnitude larger than
the values of even $v_1$ in Pb+Pb collisions (while the magnitude of
the odd $v_1$ component at LHC is only about 3 times smaller than that
at top RHIC energies) - which is consistent with much stronger values
of $\mean{\cos(\Psi_{13}-\Psi_{sp})}$ in Cu+Au collisions compared to
Pb+Pb collisions as shown in Fig.~\ref{fig:cosPsi13}.

In summary, we have analyzed the recent directed flow measurements at
midrapidity in Pb+Pb collisions at LHC and Cu+Au collisions at RHIC
in order to determine the direction of flow of the spectator
nucleons. We conclude that all the measurements strongly supports the
picture of spectators flowing ``outward'' from the collision
center-line.

\vspace{1mm}
\section{acknowledgements}
The authors appreciate early discussion of this question with
Dr. I.~Selyuzhenkov and thank Dr. C.~Ogilvie for providing references
to low energy measurements. 
This material is based upon work supported by the U.S. Department of 
Energy Office of Science, Office of Nuclear Physics under Award 
Number DE-FG02-92ER-40713.


\end{document}